\documentstyle[12pt]{article}
\advance\textheight by 60pt
\advance\voffset by -40pt
\advance\textwidth by 50pt
\advance\oddsidemargin by -25pt
\advance\evensidemargin by -25pt

\def\single_space{\baselineskip 12pt plus 1pt minus 1pt}
\def\one_and_a_half_space{\baselineskip 19pt plus 1pt minus 1pt}
\def\double_spacesp{\baselineskip 25pt plus 2pt minus 2pt}
\begin{document}
\begin{titlepage}
\begin{flushright}
{\bf
PSU/TH/113\\
MAD/PH/709\\
August 1992\\
}
\end{flushright}
\vskip 1.5cm
{\Large
{\bf
\begin{center}
Spin-Spin Asymmetries in Large Transverse \\
Momentum Higgs Boson Production \\
\end{center}
}}
\vskip 1.0cm
\begin{center}
M. A. Doncheski\\
Department of Physics\\
University of Wisconsin\\
Madison, WI 53706\\
\vskip .3cm
and\\
\vskip .3cm
R. W. Robinett and L. Weinkauf \\
Department of Physics\\
The Pennsylvania State University\\
University Park, PA 16802 \\
\end{center}
\vskip 1.0cm
\begin{abstract}

We examine the spin-dependence of standard model Higgs boson production at
large transverse momentum via the processes $gg \rightarrow gH^0$,
$qg \rightarrow qH^0$, and $q\overline{q} \rightarrow gH^0$.  The partonic
level spin-spin asymmetries ($\hat{a}_{LL}$) for these processes are large at
SSC/LHC energies.
\end{abstract}
\end{titlepage}
\double_spacesp

The prospects for probing the spin-dependence of the standard model of
particle physics at collider and supercollider energies have recently
received renewed attention \cite{workshop}, partly because of the successful
experimental tests of the Siberian snake concept \cite{krisch}.  Polarization
options at collider energies\cite{sofferrhic,particleworld} and supercollider
energies \cite{courant,soffer} and the physics programs possible at such
facilities have been discussed extensively.

Predictions for the longitudinal spin structure of hard scattering events at
such energies require two ingredients: a knowledge of the helicity structure of
the contributing matrix elements and parameterizations of the spin-dependent
parton distributions of the proton.  Lowest order predictions for the spin
structure of many standard collider processes now exist, often quoted as
partonic level asymmetries,
\begin{displaymath}
\hat{a}_{LL} \equiv
\frac{\hat{\sigma}(++)-\hat{\sigma}(+-)}
{\hat{\sigma}(++)+\hat{\sigma}(+-)}  \;,
\end{displaymath}
(where $\pm$ refers to the helicity of the incident partons parallel or
antiparallel to the polarized incident nucleon.)  Polarized lepton-nucleon
production experiments have provided some knowledge of the spin structure of
the valence quark distribution in the nucleon (at relatively large $x$) but
the spin dependence of the sea quark and gluon distributions is yet to be
measured directly, and, in fact, a large part of a program of polarized $pp$
collisions at RHIC would be a comprehensive mapping out of these
distributions\cite{loi}.

As mentioned above, many calculations of the partonic level asymmetries for
standard model processes exist (see, e.g.,
\cite{soffer},\,\,\,\cite{doncheski}--\cite{2photon} and references
therein).  While likely not relevant for a first generation program of
polarized $pp$ collisions at collider energies, the production of the
standard model Higgs boson  ($H^0$) at a polarized supercollider (such as SSC
or LHC) has also been discussed to some extent \cite{soffer}.  The
calculations of Ref. \cite{soffer} have focused on low transverse momentum
$H^0$ production via gluon fusion ($gg \rightarrow H^0$) and weak gauge boson
fusion ($qq qq\rightarrow W^+W^- \rightarrow H^0$) and the partonic level
asymmetries in these cases are known to be quite large, e.g.
$\hat{a}_{LL}(gg \rightarrow H^0) = +1$.

Higgs boson production at large transverse momentum, via the processes
$gg \rightarrow gH^0$, $qg \rightarrow qH^0$, and
$q\overline{q} \rightarrow gH^0$ has also been considered in the unpolarized
case by Ellis {\it et al.} \cite{ellis} and Baur and Glover \cite{baur}.  The
matrix elements for these processes, depending as they do on heavy quark
loops, are, in general, quite complicated functions of the kinematic
variables $\hat{s}$, $\hat{t}$, $\hat{u}$, $M_t$, and $M_H$ but simplify
dramatically in the limit of $M_t >> M_H, \hat{s}, \hat{u}, \hat{t}$. It was
noted in Ref. \cite{2photon}, that in this limit the partonic level
asymmetries for these processes reduce to those for the production of a
$^1S_0$ quarkonium state, which are known to be reasonably large\cite{rick}.
In this note, we extend this preliminary observation concerning the spin
dependence of large transverse momentum Higgs boson production in three
obvious ways.  We first explore the effects of a finite heavy quark mass on
the partonic level asymmetries ($\hat{a}_{LL}$) and then calculate the
average spin-spin asymmetry $<\hat{a}_{LL}>$ in high $p_T$ $H^0$ production
at SSC and LHC energies to determine a rough measure of the `analyzing power'
of such processes.  Finally, we estimate the range in observable longitudinal
spin-spin asymmetries $A_{LL}$ by using several parameterizations of the
polarized gluon distributions indicating a range in the current uncertainty
about the contribution of gluons to the proton spin.

The cross-section for $q \overline{q} \rightarrow H^0 g$ (and the crossed
process $qg \rightarrow q H^0$) depend only on a single heavy quark triangle
graph and the resulting $ggH^0$ form factor appears as a simple
multiplicative factor in the amplitude.  While this factor changes the total
cross-section, it as no effect on the helicity structure of the matrix
elements so the partonic level spin-spin asymmetries for these two processes
are independent of $M_t$ and are given by
\begin{equation}
\hat{a}_{LL}(qg \rightarrow qH^0) =
\frac{\hat{s}^2-\hat{u}^2}{\hat{s}^2+\hat{u}^2}
\end{equation}
and
\begin{equation}
\hat{a}_{LL}(q\overline{q} \rightarrow gH^0) = -1.
\end{equation}
These asymmetries were plotted for several values of $\sqrt{\hat{s}}/M_H$
in Ref. \cite{2photon} for illustration.  The amplitude for the dominant
$gg \rightarrow g H^0$ process depends on both triangle and box diagrams and
can be expanded in terms of two invariant functions (see, e.g.,
Ref. \cite{ellis}, Eqn.(A.6)) so that the helicity structure of the
interaction does, in fact, depend on the mass of the quark in the internal
loops.  In the notation of Ref. \cite{ellis}, one has
\begin{equation}
\frac{d\hat{\sigma}}{d\hat{t}} =
\frac{1}{16\pi\hat{s}^2}\frac{1}{4\!\cdot\! 64}
\sum_{\mbox{spins,colors}}|{\cal M}|^2
\end{equation}
where the spin and color summed invariant matrix elements are given by
\begin{eqnarray}
\sum|{\cal M}|^2 & = &
\alpha_w\alpha_S^3\frac{96}{\hat{s}\hat{t}\hat{u}}
\frac{M_H^8}{M_W^2}
(|A_2(\hat{s},\hat{t},\hat{u})|^2
+|A_2(\hat{u},\hat{s},\hat{t})|^2 \nonumber \\
& &
\;\;\;\;\;\;\;\;\;\;\;\;\;\;\;\;\;\;\;\;\;\;
+|A_2(\hat{t},\hat{u},\hat{s})|^2
   +|A_4(\hat{s},\hat{t},\hat{u})|^2 )\;.
\end{eqnarray}
The dimensionless functions $A_2$ and $A_4$ are given in terms of standard
loop integrals and thus depend on the quark loop mass and are actually
proportional to the gluon helicity amplitudes which we require for the
partonic level spin-spin asymmetry.  In fact, we find
\begin{equation}
\hat{a}_{LL}(gg\rightarrow gH^0) =
\frac{
|A_4(\hat{s},\hat{t},\hat{u})|^2
+|A_2(\hat{s},\hat{t},\hat{u})|^2
-|A_2(\hat{u},\hat{s},\hat{t})|^2
-|A_2(\hat{t},\hat{u},\hat{s})|^2}
{|A_4(\hat{s},\hat{t},\hat{u})|^2
+|A_2(\hat{s},\hat{t},\hat{u})|^2
+|A_2(\hat{u},\hat{s},\hat{t})|^2
+|A_2(\hat{t},\hat{u},\hat{s})|^2}.
\end{equation}
In the $M_t \rightarrow \infty$ limit, one has
\begin{equation}
A_4(\hat{s},\hat{t},\hat{u}) \rightarrow -\frac{1}{3}
\;\;\;\;  ,  \;\;\;\;
A_2(\hat{s},\hat{t},\hat{u}) \rightarrow -\frac{\hat{s}^2}{3M_H^4}
\end{equation}
so that the partonic level asymmetry is simply
\begin{equation}
\hat{a}_{LL}(gg\rightarrow gH^0) =
\frac{M_H^8 + \hat{s}^4 -\hat{t}^4 -\hat{u}^4}
     {M_H^8 + \hat{s}^4 +\hat{t}^4 +\hat{u}^4}
\end{equation}
which is the same result as for $gg \rightarrow g\,^1S_0$ quarkonium
production as found in Ref. \cite{2photon}.

As the only change is in the purely gluon induced processes, we illustrate
results for that sector only.  In Fig. 1 we plot the partonic level
asymmetries versus the center-of-mass angle ($\cos(\theta^*)$) for several
values of $\hat{s}/M_H^2$ in the large $M_t$ limit.  (We note that when
$\hat{s} \rightarrow M_H^2$ the asymmetry approaches the limit $+1$,
corresponding to $gg \rightarrow H^0$, independent of angle.)  In Fig. 2, we
then plot the {\it ratio} of the `exact' expressions for the partonic level
$\hat{a}_{LL}$ (Eqn. 5) to the infinite quark mass limit (Eqn. 7) versus
$\cos(\theta^*)$ for the same values of $\hat{s}/M_H^2$ and two values of
$M_t/M_H$.  In general, the asymmetries are somewhat reduced for finite
values of $M_t$ but not dramatically so.

We next plot in Fig. 3 the differential cross-sections for large transverse
momentum $H^0$ production, $d\sigma /dp_T$ versus $p_T$\,
 for two choices of
$M_H$ and for SSC and LHC energies to acquire some feel for the event rates
possible using these mechanisms.  As a test of our calculation, we are able
to reproduce the corresponding figures from Refs. \cite{ellis,baur}.  Then
to ensure that the spin-dependence of the matrix elements is large in the
kinematic regions for high $p_T$ Higgs boson production, in Fig. 4 we plot
the average partonic level asymmetry, defined via
\begin{equation}
<\hat{a}_{LL}> \equiv
\frac{\sum_{ij}
\int dx_a \int dx_b f_i(x_a,Q^2) f_j(x_b,Q^2)
\;d\hat{\sigma}_{ij}\cdot \hat{a}_{LL}^{ij}}
{\sum_{ij}
\int dx_a \int dx_b f_i(x_a,Q^2) f_j(x_b,Q^2)
\;d\hat{\sigma}_{ij} }
\end{equation}
where $f_i(x,Q^2)$ are the appropriate parton distributions.  We use EHLQ2
distributions \cite{ehlq} for consistency with Ref. \cite{ellis} as well as
the choice of momentum scale $Q^2 = M_H^2 + p_T^2$, and include all three
relevant subprocesses.  We see that the average partonic level asymmetries
are quite reasonable in all of the kinematic regimes relevant for high $p_T$
Higgs production.  Finally, we can include the effects of the polarized
parton distributions by calculating values of the observable spin-spin
asymmetry,
\begin{equation}
A_{LL} \equiv
\frac{\sum_{ij}
\int dx_a \int dx_b \Delta f_i(x_a,Q^2) \Delta f_j(x_b,Q^2)
\;d\hat{\sigma}_{ij}\cdot \hat{a}_{LL}^{ij}}
{\sum_{ij}
\int dx_a \int dx_b f_i(x_a,Q^2) f_j(x_b,Q^2)
\;d\hat{\sigma}_{ij}}
\end{equation}
where now the $\Delta f_i(x,Q^2) \equiv f_i^{(+)}(x,Q^2)- f_i^{(-)}(x,Q^2)$
are defined in terms of the spin dependent parton distributions.  Using the
parameterizations of the polarized parton distributions given by Soffer
{\it et al.} \cite{soffer} and Bourrely {\it et al.} \cite{bourrely}, we then
find the asymmetries plotted in Fig. 5.  These two choices are representative
of a standardly small gluon polarization in the nucleon (Ref. \cite{soffer})
and EMC-motivated `large' gluon contribution to the proton spin
(Ref. \cite{bourrely}) and therefore give some idea of the possible range in
observable spin dependence in this reaction.  In neither case is the
observable asymmetry larger that $1\%$; since the average partonic level
asymmetry, $< \hat{a}_{LL} >$, was seen to be large (Fig. 4), the small
observable asymmetry is due to the smallness of the polarized gluon
distribution in the kinematic region probed.

In conclusion, we have found that the partonic level longitudinal spin-spin
asymmetries in the kinematic regions relevant for large $p_T$ Higgs boson
production are quite large but that the observable spin dependence depends
critically on the, as yet unmeasured, polarized gluon distribution in the
proton.  Hopefully, measurements of various standard model processes (such as
jet and direct photon production) with a polarized proton-proton colliding
beam facility such as at RHIC will provide the first direct information on
the size of the gluon contribution to the proton spin.

We are very grateful for conversations with U. Baur and R. Stuart, who also
kindly provided us with various computer programs.  This work was supported in
part by the National Science Foundation under grant PHY--9001744 (R.R.), by
the Texas National Research Laboratory Commission under an SSC Junior Faculty
Fellowship (R.R.), by the University of Wisconsin Alumni Research Foundation
(M.D.), by the U. S. Department of Energy under contract DE-AC02-76ER00881
(M.D.), and by the Texas National Research Laboratory Commission under grant
No. RGFY9173 (M.D.).
\vskip 1cm

\newpage


%
\newpage
{\Large
Figure Captions}
\begin{itemize}
\item[Fig.\thinspace 1.] Partonic level asymmetries $\hat{a}_{LL}$ for
$gg \rightarrow gH^0$ versus $\cos(\theta^*)$ (where $\theta^*$ is the
center-of-mass scattering angle) in the $M_t \rightarrow \infty$ limit.
$\hat{s}/M_H^2 = 2\,(20,\,200)$ is shown in the solid (dashed, dotdashed)
curve.
\item[Fig.\thinspace 2.] Ratio of `exact' partonic level asymmetry
$\hat{a}_{LL}$ to that in the $M_t \rightarrow \infty$ limit
($\hat{a}_{LL}(\infty)$) versus $\cos(\theta^*)$ for two values of $M_t/M_H$
($M_t/M_H = 0.2\;(0.8)$ on the left (right) respectively).  Three values of
$\hat{s}/M_H^2$ are shown as in Fig. 1.  We use the fact that the angular
distribution is symmetric around $y = \cos(\theta^*) = 0$.
\item[Fig.\thinspace 3.]  Differential cross-section,
$d\sigma/dp_T\;(nb/GeV)$ versus $p_T\;(GeV)$ for Higgs boson production for
$M_H = 100\, GeV,\;(200\, GeV)$ for $\sqrt{s} = 40\, TeV$ dashed (dotdashed)
curve, and for $\sqrt{s} = 17\, TeV$ solid (dotted) curve.  The parton
distributions of Ref. \cite{ehlq} are used (EHLQ2) with the scale choice
$Q^2 = M_H^2 + p_T^2$.
\item[Fig.\thinspace 4.]  The average partonic level asymmetry
$<\hat{a}_{LL}>$ (as defined in Eqn. 8) in the quantity $d\sigma/dp_T$ versus
$p_T\,(GeV)$.  Curves are labelled as in Fig. 3.
\item[Fig.\thinspace 5.]  The observable spin-spin asymmetry, $A_{LL}$
(defined in Eqn. 9) in the quantity $d\sigma/dp_T$ versus $p_T\,(GeV)$.
Asymmetries for $\sqrt{s} = 40 \;TeV$, $M_H = 200\;GeV$ and
$M_t = 130\,(\infty)\,GeV$ for the polarized parton distribution functions of
Ref. [6] solid (dotted) curve, and Ref. [15] dashed (dotdashed) curve.
\end{itemize}
\end{document}